\documentclass[useAMS,usenatbib,usegraphicx]{mn2e}

\title[The measurement errors in the {\it Swift}-UVOT and {\it XMM}-OM]
 {The measurement errors in the {\it Swift}-UVOT and {\it XMM}-OM}

\author[N.P.M. Kuin, and S.R. Rosen]
  {N.P.M. Kuin$^1$\thanks{email: npmk@mssl.ucl.ac.uk} and S.R.~Rosen$^2$\thanks{email: srr11@star.le.ac.uk}\\
$^1$Mullard Space Science Laboratory, Holmbury St. Mary, Dorking, Surrey, RH5 6NT, UK,\\
$^2$Department of Physics and Astronomy,
University of Leicester,University Road,
Leicester, LE1 7RH, UK
}

\begin{document}

\date{Accepted   Received   ; in original form ; submitted to MNRAS, 2007 September 7}
\pagerange{\pageref{firstpage}--\pageref{lastpage}}
\pubyear{2007}

\maketitle

\begin{abstract}
The probability of photon measurement in some photon counting instrumentation, 
such as the Optical Monitor on the XMM-Newton satellite, and the UVOT 
on the Swift satellite, does not follow a Poisson distribution due to 
the detector characteristics, but a Binomial distribution.  
For a single-pixel approximation, an expression was derived for the 
incident countrate as a function of the measured count rate by \citet{Fordham}.
We show that the measured countrate error is binomial, and  
extend their formalism to derive the error in the incident 
count rate. 
The error on the incident count rate at large count rates is larger 
than the Poisson-error of the incident count rate. 
\end{abstract}

\begin{keywords} 
  instrumentation: detectors -- 
  methods: statistical -- 
  techniques: photometric --
  methods: data analysis
\end{keywords}

\section{Introduction}

In recent years photon-counting detectors have come into operation
in for example the UltraViolet/Optical Telescope (UVOT; \citet{Roming}) 
on the {\it Swift} gamma-ray bursts satellite, and the {\it XMM} Optical 
Monitor (OM; \citet{Mason}). 
The MIC detectors used in these instruments have been discussed by 
\citet{Fordham}. 
These photon-counting detectors operate as follows:
Incoming photons exite electrons on a photo-cathode. 
The electrons are amplified by a stack of microchannel plates and 
then the amplified electron signal is converted back to a light-pulse 
using a phosphor screen. 
Below this, a fibre bundle directs the light to a fast-scanning, 
frame-transfer CCD. 
After each frame is read-out, the 
resulting charge events in the CCD are centroided by the on-board electronics. 
 
At high incident fluxes, a photon-counting detector is limited  
due to coincident photon arrivals in a single read-out of the detector.
This represents a clear difference between the photon-counting technique 
and measurements made by direct illumination of a CCD, which can handle 
large fluxes, but has a higher background.  

Normally, when measuring the number of counts arriving in a certain 
time interval, little futher thought is given to the statistics of 
such a measurement, which were worked out long ago by \citet{Poisson}. 
Indeed, photon counting instrumentation, like photo-multiplier tubes, are 
usually seen as an exemplary case of Poisson statistics. 
However, due to the instrumental limitations imposed by centroiding 
and event-detection of the MIC detectors, no 
more than a single event recording per pixel is possible in the smallest 
timeslice of measurement. 
This handicap prevents the full distribution 
of photon arrivals being sampled and thus the measurements are not 
Poissonian, though the incoming photons follow a 
Poissonian distribution. 
As a result the errors on the photometry from the UVOT and 
OM do not follow Poisson statistics. 

For each observation, however, one can derive the measurement statistics, 
which we show in section \ref{sec2} to follow a 
Binomial distribution, and relate 
them to the Poisson distribution of the incident photons.  
Based on the measured distribution and the functional relation 
that it has to the incident Poisson distribution, 
we derive the errors in the measurement and in the inferred 
incident photon count rate in section \ref{sec2}. This paper 
aims at providing the users of the UVOT, OM and similar instruments, 
a proper way to estimate the errors in their photometry.

\section{Theory}
\label{sec2}

\subsection{The mean number of incoming photons related to the measured count}

For the detectors of interest, an exposure will be for a certain time period
$\Delta T$ and consist of $N_f$ time-slices usually called `frames'. 
Exposing and reading out each frame takes a certain fixed 
time $T_f = {\Delta T}/N_f$, called the frame-time. Since during
read-out of the detector no incoming photons are detected, a 
fraction $f_d$, called the dead-time, needs to be accounted for when 
determining the count rate. 
\footnote{This is a simplification, since during the frame-transfer time, photons
arrive, and charge is deposited in the CCD, they can be centroided 
into events when bright enough. The charge shunting process during frame 
transfer does lead to charge from a star (a fixed-position source) being 
'smeared' out and this leads to read-out streaks from bright stars.}

In the following, we will use variables for the total observation. For example, 
observed counts refer to all observed counts during the observation. This 
simplifies the treatment of the errors somewhat, and conversion to commonly used
count rates and their errors is quite straightforward. 

Now consider a single pixel. During an exposure $N_f$ measurements are taken 
from that pixel, measuring either 0 or 1 count per frame, since coincident counts 
are recorded as a single event.  It is here, where 
the difference with a Poissonian measurement comes in, since multiple detections  
in a single frame count only for one. We can use that fact to relate the probability 
of observing 1 or 0 photons to the fact that the incoming photons follow a 
Poisson distribution. 

The Poisson probability that $k$ incoming photons 
fall on one frame is a function of the mean 
incident counts per frame $\umu$:
\begin{equation}
   P(k,\umu) = \frac{e^{-\umu}\umu^k}{k!}
\end{equation}
The first two moments of the Poisson distribution are  
$\sum_{k=0}^{\infty} {P(k,\umu) = 1}$ 
and 
$\sum_{k=0}^{\infty} {k P(k,\umu) = \umu}$. 
The effective exposure time is less than the elapsed time due to the dead time. 
Therefore, the mean number of incoming photons $C_i$ during the observation 
relates to the mean probability of measurement as $\umu = C_i (1 - f_d) / N_f = \alpha C_i / N_f $, 
where $\alpha = (1 - f_d$) has been introduced for notational convenience.

The measured number of photons in $N_f$ frames, 
considering that for $k > 1$ only one photon is counted, is 
\begin{equation}
   C_o = N_f [ 0.P(0,\umu) + 1.P(1,\umu) + 1.(P(2,\umu) ... ] 
\end{equation}
Using the equations above, this can be written as
\begin{equation}
\label{eq1}
   C_o/N_f = 1 - e^{-\umu} = 1 - e^{-\alpha C_i/N_f}
\end{equation}

This functionally relates the incoming counts to the measured counts, and was originally 
derived by \citet {Fordham}.

\subsection{The error in the measured counts}

We first show that the incident Poisson distribution leads to an observed 
binomial distribution due to the coincidence-loss in the measurements, 
and then discuss the calculation of the measurement errors.

If we had an instrument that would be able to record the incoming photon distribution, 
the probability of recording $m$ incident photons in $N_f$ frames is given by the 
Poisson distribution. 
In actuality, not more than one photon can be measured per frame, so the 
distribution becomes modified in that term. Therefore, 
the probability of recording $m$ incident photons in $N_f$ frames is given by: 
\begin{equation}
  \hat{P}(k;N_f,\umu) = {N_f\choose k} P(0,\umu)^{(N_f-k)} P(m>1;\umu)^k.
\end{equation}
Where $m$ reduces to $k$ measured photons, since for each frame where $m > 1$, 
only one count is recorded.  

Substituting $k$ for $m$, using  equation \ref{eq1}, and defining 
for convience $p=e^{-\mu}$ we can rewrite this as:
\begin{equation}
  \hat{P}(k;N_f,p) = {N_f\choose k} p^{(N_f-k)} (1-p)^k,
\end{equation}
which is indeed a Binomial distribution. That means that the observed counts are 
are governed by a Binomial distribution, and that errors need to be accounted for
accordingly.

The observed error in the mean number of counts in the observation $C_o$ 
for the Binomial measured distribution will be determined by the Binomial error  
\begin{equation}
   \sigma_o  = \sqrt{C_o(N_f-Co)/N_f}.
\end{equation}

Using the observed error, the incident photon count rate error can be derived 
using the non-linear equation \ref{eq1}, because the relation has a 1-1 correspondence. 
Substracting the mean count rate from the count rate 
with a $1\sigma$ error added or substracted, we obtain the following expression 
relating the upper and lower error $\sigma_i$ in the incident counts to the error 
in the observed counts:
\begin{equation}
\label{eq4}
 \sigma_i^{+}  = -\frac{N_f}{\alpha} ln(1 + {\frac{\sigma_o}{N_f-C_o}})
\end{equation}
\begin{equation}
 \sigma_i^{-}  = -\frac{N_f}{\alpha} ln(1 - {\frac{\sigma_o}{N_f-C_o}}).
\end{equation}

For the highest incoming photon fluxes, the upper error becomes larger than the lower error, but for 
frame rates less than 0.9, the error is in a linear regime and they are nearly equal in absolute size.

\begin{figure}

\includegraphics[width=59mm,angle=90]{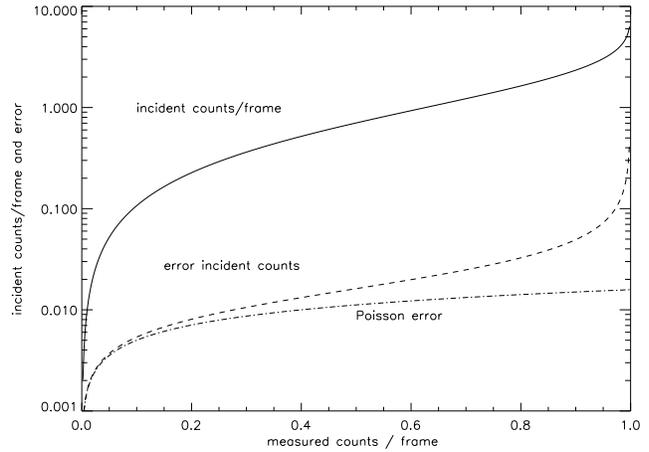}

\caption{
The ratio of the incident count rate normalised to the frame time 
is shown as a function of counts per frame (=count rate/frame rate) 
along with its error (dashed), see Eq. \ref{eq4}. For comparison, 
the error in the Poisson-limit  has been plotted also. The assumed 
number of frames for error computation was 4 000.
}

\label{fig1}

\end{figure}

\section{Discussion}

\subsection{Asymmetry}

For a point source with a certain count rate, the incoming counts will 
fluctuate in a Poissonian sense around the mean. As discussed in section \ref{sec2}, 
the measured counts are binomial. Because of this, the counts above the mean
will be mapped into a smaller range of observed count rate than those below the mean, 
which is ultimately due to the coincidence-loss.  In this sense, the width 
of the distribution, as defined by $\sigma_o$ is not an equal measure for the 
area under the distribution above and below the mean. We therefore need to 
be careful when interpreting the standard deviation derived here, especially 
for high observed counts per frame values. 

\subsection{Mapping of the uncertainty range}

There is a certain inherent width in the distribution of incoming counts which 
results in Poissonian variation around the mean, usually expressed as the Poisson 
error. The question is how that error relates to the final error in the measurement. 

\begin{figure}

\includegraphics[width=78mm]{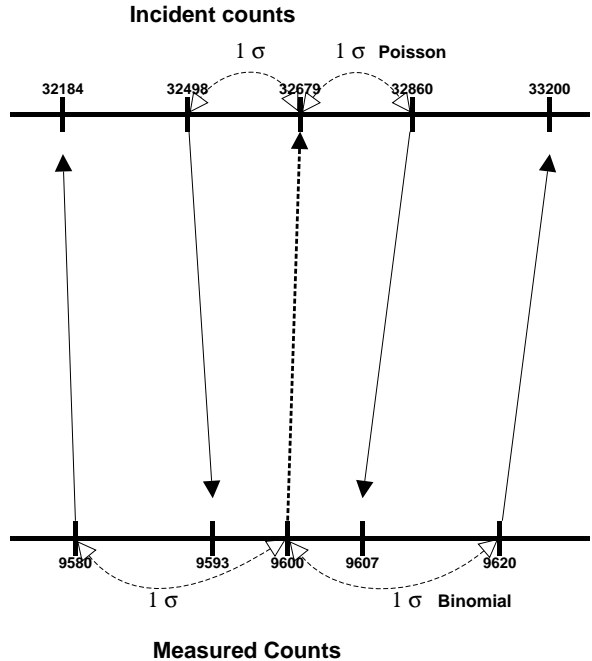}

\caption{Schematic illustration of the effect of the coincidence-loss on the errors. The top line
represents the incident counts, the bottom line the measured counts. The error
in the incident counts due to Poisson noise is indicated and how it projects to
the measured counts. The Binomial error on the measured counts has been 
indicated and how it maps to the incident counts.
}
\label{fig2}

\end{figure}

In the limit of a small number of counts per frame, they become equal.  
For larger numbers of counts per frame they diverge, and the  
measurement error, after being mapped back to the uncertainty range in the incoming 
count rate, becomes dominant. Since the magnitude of this effect is not very apparent
from the theory above, an example has been prepared in Figure \ref{fig2}.

For simplicity, the number of observed counts has been set at $C_o = 9600$ for $N_f = 10~000$
frames. The dead-time is assumed to give $\alpha = 0.985$. Using the equations above, 
the incident rate is then $C_i = 32~679$, with an associated Poisson error of 181 counts. 
In the figure we place the incident counts and its error on the top horizontal line. 
If we map the incident counts at $\pm 1 \sigma$ to the measured values they come out to
be 7 counts above and below the mean observed counts. The $1\sigma$ Binomial error on the 
observed counts, however, is 20, much larger than what would be expected from the 
mapped-back incident distribution. Mapping the measured counts at $\pm 1\sigma $ from the 
measured counts back to the incoming counts, it is readily seen that these have a
much larger spread than the incoming distribution. This effect becomes smaller for
lower ratios of $C_o/N_f$. Please note that the values we chose for our example 
have a high coincidence-loss which makes these effects more discernable.

\subsection{Confidence levels}

Confidence levels measure what percentage of the distribution of the 
measured quantity fall within certain limits. In a way they are more useful 
than the standard deviation in the presence of asymmetries, because they 
provide information on the reliability of the measurement.  
It is well known how to  determine confidence levels for the 
measured count rate, because it follows the well-known binomial distribution 
\citep{Gehrels}. However, the values
reported are the incident count rates which bear a non-linear relation
to the measured ones. Likewise, a certain confidence level in the measured
count rate will not imply the same level in the incident count rate, 
precisely because of the asymmetry mentioned above. The effect is largest
at the highest count rates, where we showed by example above, that 
the measured distribution is much broader than the incident (Poisson)
distribution.  As a result, at high count per frame rates,  
the uncertainties in the measured count rate 
dominate those in the derived incident count rate.
Also, in the limit of a low 
number of counts per frame the binomial 
confidence levels on the measured counts 
will approach the confidence levels of the 
Poisson-distributed incident counts because the distributions are 
identical in the low limit. The coincidence-loss correction at the 
limit of low counts is also negligible. 
This suggests that using the confidence limits for the measured binomial 
counts will be a good approximation for the confidence limits on the 
derived incident count rate.. 

\subsection{Background}

In general, for low count rates the effects from coincidence-loss are negligible. 
This is especially true for the background. However, it was found that in some UVOT 
observations a correction for coincidence-loss to the background was necessary and 
had an impact on the net source rates derived. 
Since the background is diffuse in nature, the arguments brought forward 
for considering the coincidence-loss in diffuse situations by  \citet{Fordham}
need to be taken into account. They discussed this case in terms of the coincidence-loss
area over which coincidence-loss acts and the exposure area. Their equation  
reverts to the single pixel case for the background. 

It is therefore important to realize that the expressions above, which were derived 
in the single-pixel approximation, need to be applied with caution to the background. 
If the measurement background area covers more than one CCD pixel, a normalization to
the coincidence area, which is presumably 
one CCD pixel, needs to be made to apply the formulas above. For example, if 
a physical pixel has 8x8 subpixels, the normalisation is as follows. 
If $C_B$ background counts were measured  from a region of $X$ subpixels, 
$X$ larger than 64 subpixels,
then the coincidence-loss correction for the background should be based on
$64 C_B / X$ counts. In practice, the correction is not as firmly known as that because
the centroiding may make the coincidence area larger or smaller. The UVOT ftools software 
uses 78 subpixels which was chosen because that is close to the theoretical value and
also the pixel-area that was used to derive the empirical 
coincidence-loss correction (see \ref{sec3.4}).

\subsection{The single-pixel approximation}
\label{sec3.4}

The coincidence-loss formula under the single-pixel approximation 
has been very successful
in predicting the correct rates in the UVOT \citep{Poole}.
Other support for the use of the single-pixel-approximation 
to calculate the coincidence-loss effect on the observed count 
rate comes from studies during the construction of the detectors, 
\citep{Fordham} and the implementation of the centroiding \citep{Hajime}.  
The measurement algorithm locates the centre of the photon splash, 
which generally falls across 2-3 CCD pixels, and has an accuracy of a small 
fraction of a CCD pixel, (allowing recording of UVOT and OM data with an accuracy
of 1/8th of the physical CCD pixel size.) 
Anomalies are rejected using four out of nine CCD pixels. 
As a result, the action of coincident photons is distributed over several pixels 
on the detector and are also folded through a screening algorithm. The net effect 
turns out to be a strengthening of the single pixel approximation, although the 
exact size of the coincidence-loss region, and its relation to the physical CCD 
pixels, is still under study.    
Were the detections really independent  single-pixel measurements, then it is easy 
to show, that photon splashes which would fall in different ways over pixel-boundaries
would reduce the effects of coincidence-loss by 10-20\% at high count rates. 
In reality, a small upwards empirical correction of the order of 6\% is found 
to be needed to the theoretical 
single-pixel-rate in the UVOT \citep{Poole} and OM, which is perhaps due to loss of some 
measurements of truly 
coincident, but slightly displaced, photons. Those could 
distort the symmetry of the electron splash on the detector 
suffiently to be screened out as bad data.  

\subsection{Dead-time accounting}

In the original formulation of the coincidence-loss correction \citet{Fordham} the effects of 
the detector dead-time in each frame were discussed but were not explicitly included in the 
coincidence-loss correction equation. As a result, early corrections for the coincidence-loss
did not include this term. Since the current formulation includes this term, no further correction
for dead-time is needed after application of equation \ref{eq1}.  

\subsection{Photometric packages}

Currently most astronomical photometry software, like IRAF and DAOPHOT 
may incorrectly report  the error for measurements like these, 
because generally the assumption is made 
that the photometric measurements are dominated by Poisson-noise. 
That is considered a good assumption for photo-multiplier and normal CCD measurements. 
As we show in figure~\ref{fig2}, the Poisson measurement error underestimates 
the error in these photon-counting instruments affected by coincidence-loss.

\section{Conclusions}

We have shown in this paper how to derive the error in measurements made 
with photon-counting detectors of the type used in the {\it Swift} UVOT and {\it XMM}~OM
instruments. By comparing to the Poisson error usually used in photometry
we make clear how significant this effect can be, and consider that 
users of these instrument must use our formalism to derive the errors 
in their measurements.

\section*{acknowledgements}

We benefitted from stimulating discussions with Alice Breeveld, Tracey Poole, 
Wayne Landsman, Chris Brindle, Keith Mason, Antonio Talavera, 
and Vladimir Yershov during the development of these ideas. We thank 
Patricia Schady for comments on an early version of this paper.
Support of this work was through the Swift Operations at UCL-MSSL through a grant 
from the UK Science and Facilities Council.

\bibliographystyle{mn}

\bsp

\label{lastpage}

\end{document}